\documentclass[sigconf]{acmart}

\AtBeginDocument{%
  }

\setcopyright{none}
\renewcommand\footnotetextcopyrightpermission[1]{}
\acmConference[DAC '26]{Design Automation Conference}{July 26--29,
  2026}{Long Beach, CA}

\usepackage[T1]{fontenc}

\usepackage[dvipsnames]{xcolor}
\usepackage{hyperref}
\hypersetup{
  breaklinks=true,
  colorlinks=true,
  allcolors=BlueViolet,
}

\usepackage{graphicx}
\graphicspath{{./figures/}} %

\usepackage[yyyymmdd]{datetime} %

\usepackage{physics}
\usepackage{braket}

\usepackage{caption}
\usepackage{subcaption}
\usepackage[nameinlink,capitalize]{cleveref}
\usepackage{tikz}
\usetikzlibrary{quantikz2}
\usepackage{adjustbox}
\usepackage{multirow}
\usepackage{bm}

\usepackage{orcidlink}

\begin{document}
\title{Compilation Pipeline for Predicting Algorithmic Break-Even in an Early-Fault-Tolerant Surface Code Architecture}

\author{Tianyi Hao, Joseph Sullivan, Sivaprasad Omanakuttan, Michael A. Perlin, and Ruslan Shaydulin}
\affiliation{%
  \institution{Global Technology Applied Research, JPMorganChase}
  \city{New York}
  \state{NY}
  \country{USA}
}
\authornote{Correspondence should be addressed to \texttt{joe.sullivan@jpmchase.com}}

\begin{abstract}
Recent experimental progress in realizing surface code on hardware, including demonstrations of break-even logical memory on devices with
up to hundreds of
physical qubits, has materially advanced the prospects for fault-tolerant quantum computation. 
This progress creates urgency for the development of compilation workflows that directly target the forthcoming generation of devices with
thousands of 
physical qubits, for which algorithm execution becomes practical.
We develop a pipeline for compiling logical algorithms to physical circuits implementing lattice surgery on the surface code, and use this pipeline to identify the requirements for achieving \emph{algorithmic} break-even---where quantum error correction improves the performance of a quantum algorithm---for two prominent quantum algorithms: the quantum approximate optimization algorithm (QAOA) and quantum phase estimation (QPE).
Our pipeline integrates several open-source software tools, and leverages recent advances in error-aware unitary gate synthesis, high-fidelity magic state production, and the calculation of correlation surfaces in the surface code.
We perform classical simulations of physical Clifford proxy circuits produced by our pipeline, and find that both 5-qubit QAOA and QPE can reach algorithmic break-even with 2517 physical qubits (surface code distance $d=11$) at physical error rates of $p=10^{-3}$, or 1737 physical qubits ($d=9$) at $p=5\times 10^{-4}$.
Our work thereby identifies conditions for achieving algorithmic break-even with near-term quantum hardware and paves the way towards an end-to-end compiler for early-fault-tolerant surface code architectures.
\end{abstract}

\maketitle
\pagestyle{plain}
\section{Introduction}
Applying quantum algorithms to large-scale problems arising in science and industry necessitates fault tolerance. 
Among the approaches to fault tolerance, the surface code \cite{Dennis2002,quant-ph/9811052,fowler2012surface} remains one of the most promising error-correcting codes for achieving universal quantum computation.
The prominence of the surface code can be attributed to a high threshold for fault tolerance (that is, lenient requirements for physical error rates), nearest-neighbor qubit connectivity, and the relative simplicity of implementing logical gates with lattice surgery \cite{horsman2012surface, litinski2019game}.
Consequently, the vast majority of detailed end-to-end resource estimates for achieving a fault-tolerant quantum advantage target surface-code-based architectures~\cite{gidney2025factor2048bitrsa, fontana2025endtoendquantumalgorithmstensor, omanakuttan2025thresholdfaulttolerantquantumadvantage}.
Recent years saw rapid progress in realizing surface code on hardware, culminating in multiple demonstrations of quantum error-corrected memory below threshold: as the distance of the surface code grows, the logical error rate is suppressed exponentially~\cite{google2023suppressing, google2025quantum, Bluvstein2023}.
The experimental progress makes urgent the need to develop compilation strategies targeting the next generation of devices with thousands of physical qubits, capable of executing quantum algorithms with a modest number of logical qubits.

However, there is an increasing recognition \cite{babbush2025grand} that current compilation techniques are ill-suited to the forthcoming era of early fault-tolerant (EFT) quantum computing.
Existing studies~\cite{litinski2019game, autobraid, watkins2024high, leblond2024realistic, silva2024multi-qubit, molavi2025dependency-aware, molavi2025generating, kan2025sparo} typically assume a fixed logical qubit layout with abundant ancilla qubits and dedicated magic state factories.
As a result, these studies focus on qubit mapping and routing by finding non-overlapping CNOT paths or compiling to Pauli-product rotations.
These assumptions are only reasonable in the distant future, when the number of physical qubits is no longer a limitation, allowing for large code distances and many logical ancilla qubits.
In contrast, surface code compilation in the EFT era needs to work under drastically different conditions.
With space at a high premium, we cannot afford dedicated magic state factory regions, and physical resources must be shared between factories, data qubits, and ancilla qubits.

In this work, we present a comprehensive pipeline for compiling quantum algorithms to lattice surgery primitives targeting early-fault-tolerant quantum devices based on 2D surface code.
Our pipeline combines several existing tools \cite{qokit, rustiq, qiskit2024, trasyn, lassynth, gidney2021stim, tqec} that use different intermediate representations that are amenable to complementary optimizations.
Our implementation leverages---and in some cases improves upon---recent innovations such as direct lattice surgery synthesis~\cite{lassynth}, in-place $Y$-basis access \cite{Gidney_Y}, magic state cultivation \cite{gidney2024magicstatecultivationgrowing}, and error-aware gate synthesis \cite{omanakuttan2025thresholdfaulttolerantquantumadvantage, trasyn}.
Taking as input a logical circuit and a logical qubit layout, our pipeline produces (among other artifacts) a complete physical circuit to implement a Clifford proxy of the logical circuit.
This Clifford proxy circuit enables classical simulations that produce realistic resource estimates and hardware specifications required to achieve algorithmic break-even, or the point at which quantum error correction improves the performance of a quantum algorithm.

We showcase our pipeline by compiling logical circuits to physical circuits that fault-tolerantly implement (Clifford proxies of) quantum phase estimation (QPE), a common subroutine in quantum algorithms, and the quantum approximate optimization algorithm (QAOA)~\cite{Hogg2000,Hogg2000search,farhi_qaoa} applied to the smallest non-trivial instance of the low-autocorrelation binary sequences (LABS) problem~\cite{Boehmer1967,Schroeder1970,shaydulin2024evidencelabs}.
We then simulate the physical circuits obtained by compiling QAOA for 5-bit LABS to identify the requirements for achieving algorithmic break-even.
Specifically, we find that both test algorithms reach break-even with 2517 physical qubits (surface code distance $d = 11$) at physical error rates of $p = 10^{-3}$, or 1737 physical qubits ($d = 9$) at $p = 5 \times 10^{-4}$.

\section{Background}
\label{sec:background}
\paragraph{Surface code}
A quantum error correction (QEC) code encodes $k$ logical qubits into $n>k$ physical qubits to protect information from hardware and control errors.
The distance of a QEC code is the minimum number of single-qubit errors that results in an undetectable logical error, and a distance-$d$ code can correct $\lfloor\frac{d-1}{2}\rfloor$ single-qubit errors.
The surface code---more specifically, the \emph{rotated} surface code that we focus on in this work---encodes every logical qubit into $d^2$ physical qubits, and thereby has $n=kd^2$ \emph{data qubits}.
This code additionally requires $n-k$ \emph{syndrome qubits} to perform error correction, and $O(kd)$ \emph{padding qubits} to perform logical operations.
All qubits of the surface code can be arranged in a two-dimensional square lattice, and every qubit only needs to interact with its four nearest neighbors, making it well-suited for solid-state architectures such as superconducting qubits.
Quantum error correction is performed by running $d$ \emph{code cycles}, where one code cycle consists of running a short circuit on the data and syndrome qubits, followed by a measurement of all syndrome qubits.

\paragraph{Lattice surgery}
Logical operations on the surface code can be realized in different ways, of which lattice surgery \cite{lattice_surgery, litinski2019game} is the leading paradigm for its qubit and time efficiency.
Lattice surgery divides the entire grid of physical qubits into rectangular patches, each of which encodes one logical qubit.
Lattice surgery uses \emph{merge} and \emph{split} operations to perform multi-qubit logical operations.
Merging occurs when two or more surface code patches are converted into one composite patch by ``promoting'' the padding qubits between the merged patches to syndrome qubits and running $d$ code cycles.
Splitting separates one large surface code patch into a collection of child patches by performing single-qubit syndrome-qubit measurements along cuts of the parent patch in one cycle, and demoting the measured syndrome qubits to padding qubits.

\begin{figure*}
    \begin{minipage}{0.27\textwidth}
        \centering
        \subcaptionsetup[figure]{skip=-100pt,slc=off,margin={0pt,0pt}}
        \begin{subfigure}{\linewidth}
            \centering
            \hfill
            \begin{minipage}{0.4\linewidth}
            \begin{adjustbox}{width=\linewidth, angle=90}
            \begin{quantikz}[row sep=0cm]
                & \ctrl{2} & \ghost{Z} \\
                \setwiretype{n} & \ghost{M_Z} & \\
                & \targ{} & \ghost{X}
            \end{quantikz}
            \end{adjustbox}
            \end{minipage}
            {\Huge =}
            \begin{minipage}{0.45\linewidth}
                \includegraphics[width=\linewidth]{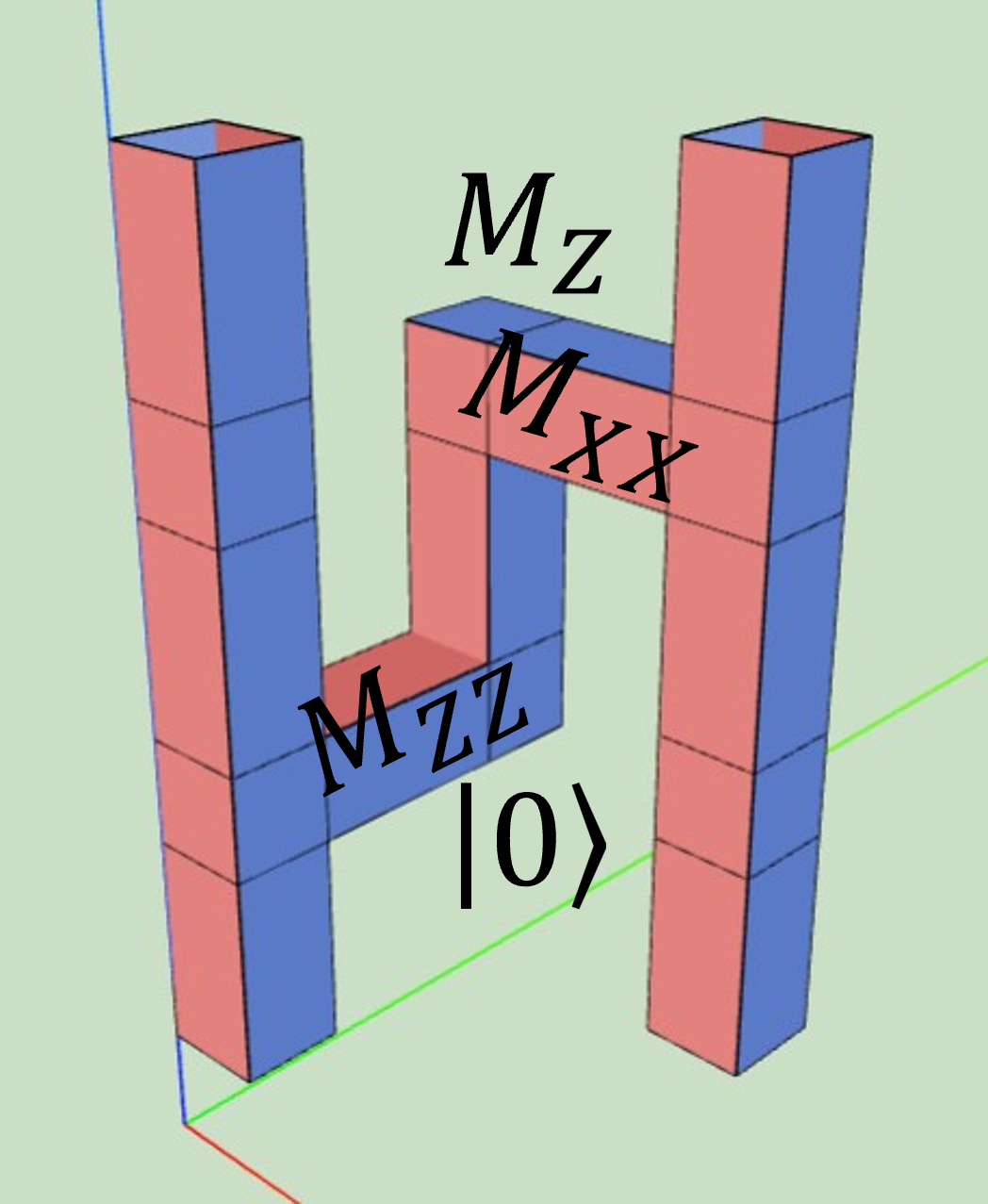}
            \end{minipage}
            \vspace{10pt}
            {\Huge \\ =}
            \begin{adjustbox}{width=0.85\linewidth}
            \begin{quantikz}[row sep=0.2cm, align equals at=2.1]
                & \gate[2]{M_{ZZ}} & \gate{Z} & &\\
                \lstick{$\ket{0}$} & & \gate[2]{M_{XX}} \wire[u]{c} & \gate{M_Z} & \setwiretype{n} \\
                & \gate{X} \wire[u]{c} & & \gate{X} \wire[u]{c} & \\
            \end{quantikz}
            \end{adjustbox}
            \subcaption{}
            \vspace{100pt}
        \end{subfigure}
        \subcaptionsetup[figure]{skip=-28pt,slc=off,margin={0pt,0pt}}
        \begin{subfigure}{\linewidth}
            \centering
            \begin{adjustbox}{width=0.8\linewidth}
            \begin{quantikz}[row sep=0.2cm]
                \lstick{$\ket{\psi}$} & \ctrl{1} & \gate{S} & \rstick{$T\ket{\psi}$}\\
                \lstick{$\ket{T}$} & \targ{} & \gate{M_Z} \wire[u]{c} & \setwiretype{n} \\
            \end{quantikz}
            \end{adjustbox}
            \subcaption{}
            \vspace{20pt}
        \end{subfigure}
        \caption{
        (a) Pipe diagram representing a CNOT gate, which directly translates to the joint Pauli measurements implementation of the CNOT. 
        The conditional Pauli gates are tracked in software.
        (b) The magic state injection circuit.}
        \label{fig:background}
    \end{minipage}
    \hfill
    \begin{minipage}{0.7\textwidth}
        \centering
        \includegraphics[width=\linewidth]{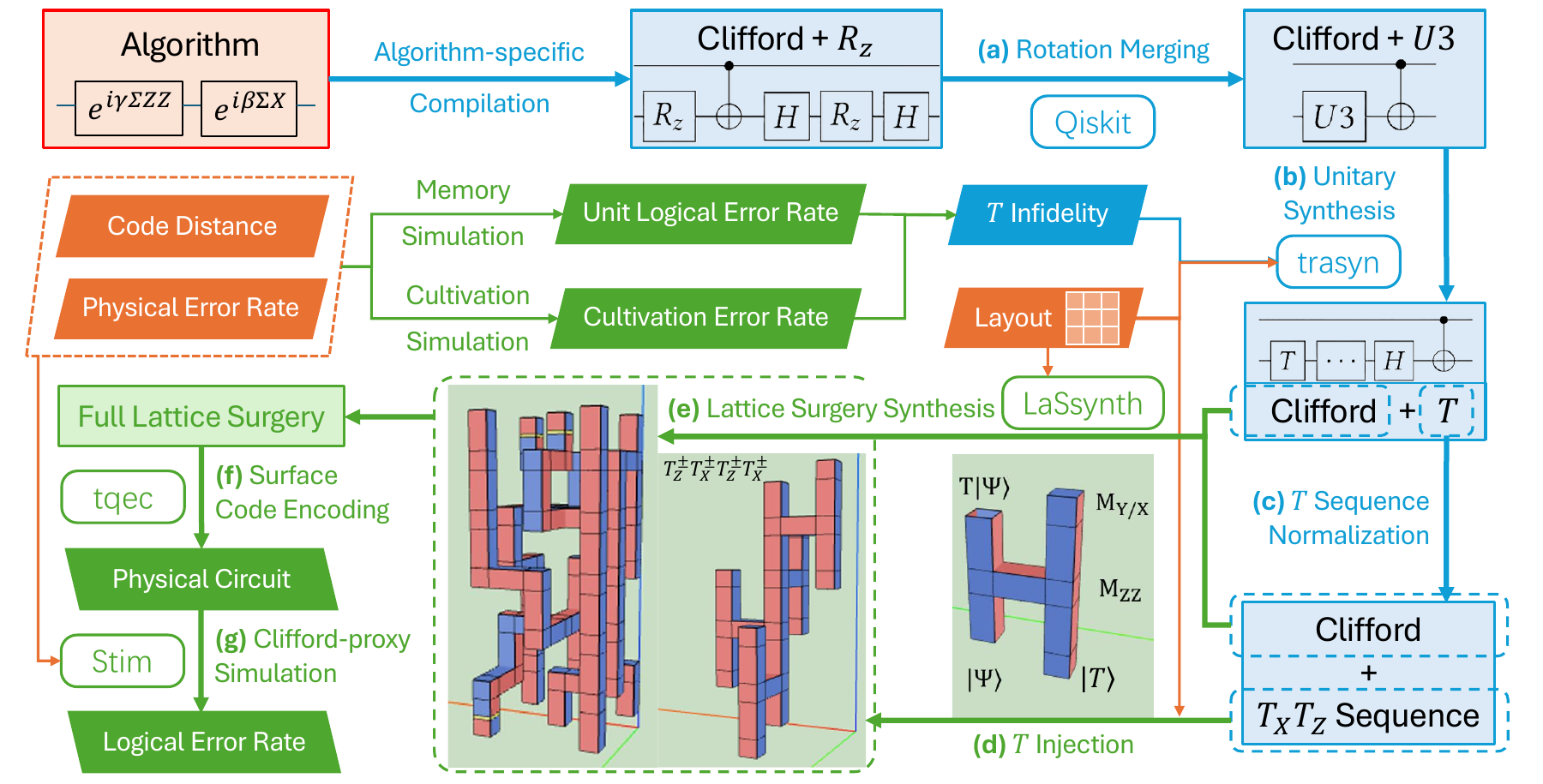}
        \captionof{figure}{The full pipeline.
        Raw inputs to the pipeline are indicated in orange.
        Intermediate representations, transformations, and metadata on the logical level (agnostic to the choice of error-correcting code) is indicated in blue, while data on the level of lattice surgery and physical qubit operations is indicated in green.
        (a) \Cref{sec:u3-transpilation}: rotation merging and $U3$ compilation. (b) \Cref{sec:rotation-synthesis}: execution-error-aware rotation synthesis. (c) \Cref{sec:t-normalization}: $T$ sequence normalization. (d) \Cref{sec:t-injection}: Magic state preparation and injection. (e) \Cref{sec:lattice-surgery-compilation}: Clifford to lattice surgery. (f) \Cref{sec:physical-circuit}: Lattice surgery to physical circuit. (g) \Cref{sec:results}: Resource estimation based on physical-level simulation.}
        \label{fig:full-pipeline}
    \end{minipage}
\end{figure*}

Lattice surgery 
can be represented diagrammatically via pipe diagrams.
\Cref{fig:background}(a) shows an example pipe diagram that implements a CNOT gate.
These diagrams have a spacetime interpretation.
A single upward column represents a surface code patch evolving in time. Each cube represents running $d$ code cycles on a logical qubit.
The total volume occupied by the cubes is referred to as the spacetime volume of the pipe diagram.
The longer pipes in between cubes do not occupy actual spacetime during computation, but help visualize the merge and split operations.
A horizontal bridge connecting columns represents a merge-split process: the bottom face and bulk of the bridge represent the $d$ cycles of a merge operation, and the top face of the bridge represents a one-cycle split operation.

\paragraph{Realization of logical operations}
At the logical level, lattice surgery is equivalent to measurement-based quantum computing.
Logical Clifford gates can be tracked in software (Pauli $X, Y, Z$), performed directly onto logical qubits ($H$), or realized by measurements ($S, CX$).
However, logical non-Clifford gates, which are essential for universal quantum computation, require special treatment.

Continuous single-qubit rotations cannot be implemented directly and fault-tolerantly in the surface code.
Instead, they need to be approximated by a sequence of discrete gates, usually a combination of the Clifford $H, S$ gates and the non-Clifford $T$ gate.
This approximation is known as rotation synthesis and leads to a synthesis error for each rotation.
Moreover, a fault-tolerant $T$ gate cannot be directly applied to a logical qubit.
Instead, it needs to be teleported from prepared $\ket{T}$ states (magic states), as shown in \Cref{fig:background}(b).
This process is referred to as magic state injection.
The preparation and injection of high-fidelity magic states is an essential computational primitive in surface code computing.

\section{Overview of Improvements to Surface Code Compilation}
Compiling a logical circuit into a surface-code-encoded physical circuit is a multi-step process involving many intermediate representations.
Below, we highlight our improvements over past studies for each subproblem encountered in this process.

\paragraph{Rotation gate synthesis}
Past fault-tolerant compilation and resource estimation studies have widely adopted an $R_Z$-based workflow, where the logical circuit is first transpiled into Clifford+$R_Z$, and each $R_Z$ is in turn synthesized into Clifford+$T$ \cite{omanakuttan2025thresholdfaulttolerantquantumadvantage, resch2021benchmarking, blunt2024compilation, stein2024architectures, watkins2024high, wang2024optimizing}.
We improve this step by adopting a $U3$-based workflow, which leads to fewer $T$ gates at the same level of synthesis error (\Cref{sec:u3-transpilation}).

\paragraph{Error budgeting}
Past studies~\cite{beverland2022AzureQRE, leblond2024realistic, fontana2025endtoendquantumalgorithmstensor, huggins2025fluid}
often set a target error for an algorithm, and allocate error budgets top-down to different parts of the execution, including synthesis errors, logical errors, and magic state preparation errors, which in turn decide the synthesis error threshold, code distance, and magic state preparation configurations.
In the EFT setting, many of these hyperparameters (such as achievable code distances and magic state errors) are determined by hardware specifications, and the compiler should aim to minimize the overall infidelity of an algorithm from the bottom up.
We implement execution-error-aware rotation synthesis, which considers the complete execution overhead of the synthesized gate sequence (including idling logical qubits), and minimizes the process infidelity incurred by the fault-tolerant execution of each rotation (\Cref{sec:rotation-synthesis}).

\paragraph{Magic state preparation and injection}
Past studies have employed magic state distillation~\cite{bravyi2005universal} for preparing magic states, a process of post-selecting many noisy input states to produce a high-fidelity output state. Due to the high spatial footprint of distillation protocols, magic states are prepared in dedicated magic state factory areas that are separate from compute qubits, which leads to an overhead of routing the prepared magic states to the injection location~\cite{autobraid, leblond2024realistic, molavi2025dependency-aware, molavi2025generating}. These works often also assume a beyond-demand production rate to ensure magic states are available whenever needed, increasing the spatial overhead. In contrast, we utilize magic state cultivation~\cite{gidney2024magicstatecultivationgrowing}, a new preparation protocol that significantly reduces the spacetime requirement. In particular, we concretely verify the practicality of an efficient consecutive injection scheme used in~\cite{gidney2025factor2048bitrsa}, which enables in-place on-demand magic state access, eliminating the routing and excessive-production overheads (\Cref{sec:t-normalization,sec:t-injection}).

\paragraph{Compilation to lattice surgery} 
Existing studies have employed Pauli-product rotations as an intermediate representation for compiling from Clifford+$T$ to lattice surgery~\cite{litinski2019game, beverland2022AzureQRE, watkins2024high, silva2024multi-qubit, kan2025sparo}. Pauli-product rotations eliminate the Clifford gates, but at the cost of incurring high-weight multi-Pauli measurements involving many qubits. High-weight measurements hinder the parallelizability of operations and make decoding more challenging, thus leading to high $T$ depths and longer execution times~\cite{beverland2022edge-disjoint, mcardle2025fast}. Furthermore, past studies assume a fixed layout of logical data and ancilla qubits, where many qubits are spent as ancillas providing access to data qubits, limiting the qubit utilization rate. In contrast, we leverage the recent development in direct lattice surgery synthesis~\cite{lassynth}, which produces lattice surgery operations based on the logic extracted from the circuit instead of performing a gate-by-gate translation. This approach does not distinguish between data and ancilla qubits and provides additional flexibility and efficiency (\Cref{sec:lattice-surgery-compilation}).

\paragraph{Resource estimation}
Instead of compiling to physical circuits, past surface code compilation studies~\cite{watkins2024high, leblond2024realistic, kan2025sparo, huggins2025fluid} stop at the lattice surgery layer. Consequently, the resource estimates are limited to the area, depth, and spacetime volume of lattice surgery operations. More crucially, fidelities of executing the target algorithms fault-tolerantly are estimated by either gate counting or volume counting, where each gate or unit volume is simply modeled with a depolarizing error. By integrating and improving several open-source tools, in particular the design automation tool \texttt{tqec}~\cite{tqec}, we assemble an automated algorithm-to-physical-circuit surface code compilation pipeline. To our best knowledge, this is the first physical-level compilation and resource estimation work for a surface-code-encoded algorithm.
Besides more accurate physical resource estimations, we also estimate logical error rates by classically simulating compiled Clifford proxy circuits with a realistic noise model, where the error propagation obeys the structure of the circuit that would be run on hardware (\Cref{sec:physical-circuit,sec:results}).

\section{An Automated Error-Adaptive Surface Code Compilation Pipeline}
\label{sec:methods}
We describe the successive steps in our pipeline that compile a quantum algorithm into a physical circuit.
The entire workflow is detailed in \Cref{fig:full-pipeline}.

\subsection{Rotation merging}
\label{sec:u3-transpilation}

Given an algorithmic circuit written in Clifford+$R_Z$, we want to synthesize the $R_Z$ gates into Clifford+$T$ while minimizing the $T$ count.
In the deterministic ancilla-free setting, an $\epsilon$-close approximation of an $R_Z$ gate (with respect to the operator norm) requires at least
$3\log_2 \frac{1}{\epsilon} + K$
$T$ gates for some constant $K$ \cite{gridsynth}.
This lower bound is nearly achieved by the widely-adopted $R_Z$ synthesizer \texttt{gridsynth}, with $K=O(\log\log1/\epsilon)$ \cite{gridsynth}.
However, the total $T$ gate count of a circuit can be reduced by compiling to Clifford+$U3$ (\Cref{fig:full-pipeline}(a)), where $U3$ is a three-parameter gate that implements an arbitrary single-qubit operation, and synthesizing $U3$ gates directly into Clifford+$T$.
Compared to Clifford+$R_Z$, Clifford+$U3$ leads to fewer rotations to synthesize, while
our $U3$-to-Clifford+$T$ decomposition incurs a similar $T$ gate cost per $U3$ gate ($\approx 3\log_2 \frac{1}{\epsilon} + 3.75$) \cite{paetznick2013repeat, PhysRevLett.109.190501} as previously required for each $R_Z$ gate.
This compilation strategy saturates a lower bound on the deterministic ancilla-free $T$ gate cost of a $U3$ gate \cite{selinger2012efficient} and is asymptotically optimal in this setting\footnote{While additional ancillas can be used to further reduce the $T$ gate cost of $R_Z$ gates~\cite{kliuchnikov2023shorter}, we focus on the more space-efficient ancilla-free setting in this work.}.
We adopt the $U3$-based workflow and synthesize $U3$ gates using \texttt{trasyn}~\cite{trasyn}. This workflow merges continuous single-qubit rotations like $R_Z$ and $R_X$ into $U3$ gates. By exploiting gate commutativity, even rotations not directly adjacent can be merged, thereby leading to fewer $T$ gates for most applications compared to the $R_Z$-based workflow~\cite{trasyn}.

\subsection{Execution-error-aware rotation synthesis}\label{sec:rotation-synthesis}

A rotation approximated by a sequence of Clifford+$T$ gates encounters two types of errors: the synthesis error from the approximation and the execution error incurred by executing the synthesized gate sequence on noisy quantum hardware.
Reducing synthesis error comes at the cost of greater $T$ gate cost, which increases execution error.
In light of this tradeoff, it has previously been observed that for a fixed noise model, there is an optimal choice of synthesis error that minimizes overall process infidelity~\cite{omanakuttan2025thresholdfaulttolerantquantumadvantage, trasyn}.
In this work, we further advance this practice by considering comprehensive execution error, which includes memory errors from idling logical qubits, in the synthesis process.

We adapt the error-biased sampling approach in \texttt{trasyn}~\cite{trasyn}. For the target rotation $U$ and the unitary $V=\prod_{g\in G} g$ from a Clifford+$T$ sequence $G$, instead of sampling based only on the synthesis fidelity $\mathrm{Tr}[U^\dagger V]$, where $\mathrm{Tr}[\cdot]$ denotes the matrix trace, we sample sequences based on the overall process fidelity 
\begin{equation}
\mathrm{Tr}\left[S^\dagger(U) \prod_{g\in G} S_{\varepsilon_g}(g)\right], 
\end{equation}
where $\varepsilon_g$ is the error channel of gate $g$ and $S_{\varepsilon_g}(\cdot)$ denotes the superoperator~\cite{wood2011tensor} such that $S\ket{\rho}\rangle=\ket{\varepsilon(\rho)}\rangle$ for arbitrary state $\rho$.
Instead of relying on a fixed global synthesis error threshold applying to all syntheses, our improvement ensures the process infidelity is individually minimized for each synthesized unitary. The general channel-based construction allows diverse error types to be considered, and we expect our execution-error-aware rotation synthesis and the bottom-up error minimization to have broader applications.

Based on this approach, we implement $T$ gates at the lattice surgery level (detailed in \Cref{sec:t-normalization,sec:lattice-surgery-compilation}), and as a simplifying approximation only apply execution errors to the $T$ gates.
There are three contributors to the $T$ execution error $e_T$: the error from the cultivation process, the error for performing the injection lattice surgery operations, and the idling logical error on all data qubits while waiting for the injection to complete.
For each code distance $d$ and physical error rate $p$ we consider, we perform physical-level surface code memory simulations to estimate $e_\mathrm{unit}$, the logical error rate for a unit spacetime volume ($d\times d\times d$). We also conduct simulations for the magic state cultivation to estimate the cultivation error $e_\mathrm{cultiv}$. We model both errors with the depolarizing model. Given our $T$ injection process (\Cref{sec:t-normalization}), for each $T$ gate we include 1.25 time units for performing the injection, $n$ units of data qubits idling, and the cultivation error $e_\mathrm{cultiv}$ to estimate the $T$ error
\begin{equation}
    e_{T} = 1 - (1 - e_\mathrm{unit})^{n+1.25}(1 - e_\mathrm{cultiv}), 
\end{equation}
which is used as a depolarizing error to perform execution-error-aware rotation syntheses, shown in \Cref{fig:full-pipeline}(b).

This process yields a unique synthesized circuit for each physical configuration $(d,p)$.
The overlap between this circuit and the logical circuit gives the circuit-level synthesis error, which we later use as a multiplicative factor in estimating the fidelity of the algorithm execution.

\subsection{Clifford+$T$ sequence normalization}\label{sec:t-normalization}
Any single-qubit Clifford+$T$ gate sequence can be transformed into a sequence of gates that is matched by the regular expression
\begin{equation}
     (H | S)? (TH | T^\dagger H)* C
\end{equation}
with the same $T$ count, where $C$ denotes an arbitrary single-qubit Clifford gate~\cite{wang2024optimizing, gidney2025factor2048bitrsa}.
This sequence can, in turn, be written in the form $T_X^\pm T_Z^\pm T_X^\pm T_Z^\pm\cdots$ up to additional Cliffords at the beginning and the end, where $T_Z^\pm=T^\pm$, $T_X^\pm=HT^\pm H$, $T^+=T$, and $T^-=T^\dagger$.
Altogether, the entire logical circuit can be written in a regular ``Clifford+$T_XT_Z$'' form (\Cref{fig:full-pipeline}(c)).
The Cliffords can be compiled directly into lattice surgery primitives (\Cref{sec:lattice-surgery-compilation}), while the $T_XT_Z$ sequences admit an efficient magic state injection procedure, detailed below.

\subsection{Magic state cultivation and injection}\label{sec:t-injection}

\begin{table}
\caption{
Cultivation statistics for $p=10^{-3}$.
}
{\small
\begin{tabular}{c ccc}
\toprule
Distance $d$ & Success rate & Logical error rate & \#Cycles/trial \\
\cmidrule(lr){1-1}\cmidrule(lr){2-4}
7 & 0.499 & $2.50 \times 10^{-5}$ & 9 \\
9 & 0.477 & $1.93 \times 10^{-5}$ & 10 \\
11 & 0.414 & $9.77 \times 10^{-6}$ & 11 \\
\bottomrule
\end{tabular}
}
\label{tab:cultivation}
\end{table}

To apply $T$ gates, we utilize an efficient magic state injection scheme proposed in~\cite{gidney2025factor2048bitrsa}, shown in \Cref{fig:full-pipeline}(d). $T_X$ injections can be performed in the same way, with the roles of $Z$ and $X$ bases switched. Compared to a CNOT-based teleportion circuit (\Cref{fig:background}(c)), this construction takes advantage of the multi-Pauli measurement, an operation natively available with lattice surgery merging and splitting, to avoid the extra ancilla required by the CNOT and reduce the cycles from $2d$ to $d$. With the advancement of the in-place $Y$-basis measurement~\cite{Gidney_Y}, the $S$ gate correction is replaced by a conditional measurement, where the measurement basis depends on the result of the previous multi-Pauli measurement. Not only does this decomposition eliminate the overhead of performing a traditional $S$ gate injection, but the dynamic part of this decomposition can be managed within a spacetime volume of $0.5d\times d \times d$ (for a $Y$-basis measurement), minimizing the impact on executing surrounding and subsequent operations.

The efficient $T_{X/Z}$ injection enables successive injections for a $T_X T_Z$ sequence, utilizing the four neighboring qubits in a round-robin (\Cref{fig:full-pipeline}(d) left). Provided that the magic state can be prepared in $2.5d\sim3d$ cycles on a single logical qubit patch, this design makes an extremely efficient use of the five qubit patches, where the center data qubit is injected every $d$ cycles, and the ancilla qubits prepare for $2.5d\sim3d$ cycles, inject for $d$ cycles, perform conditional measurement for $1\sim0.5d$ cycles, and repeat.
Below, we analyze the cost and expected fidelity of preparing magic states via cultivation in our setting.

Magic state cultivation~\cite{gidney2024magicstatecultivationgrowing} is an efficient magic state preparation protocol that builds up encoded magic states in several steps, post-selecting each time to ensure the resulting state has high fidelity. Importantly, its low spatial footprint allows it to be performed on a single surface code patch when $d\ge7$, providing in-place on-demand magic state access. Using software from Ref.~\cite{gidney2024magicstatecultivationgrowing}, we perform simulations similar to Figures 14 and 15 of Ref.~\cite{gidney2024magicstatecultivationgrowing} to confirm that the round-robin injection is feasible.
We assume SI1000, a superconducting-inspired noise model~\cite{Gidney_SI1000} parameterized by a two-qubit error rate $p$, and simulate physical circuits for surface code distances $d=7,9,11$ and $p=10^{-3}, 7\times 10^{-4}, 5\times 10^{-4}$.
In cultivation, the complementary gap is ``soft information'' from decoding that serves as a heuristic indicator of state quality. By tuning the complementary gap threshold, we can trade between cultivation success rate and logical error rate, such that a successful cultivation takes around $2.5d$ cycles.
We assume that decoding for the complementary gap can finish in two code cycles.
This assumption is consistent with existing FPGA implementations of the Union-Find decoder in Ref.~\cite{riverlane_2025}, though we note further improvements may be necessary to match the performance of the Minimum-Weight Perfect Matching~\cite{Higgott2025sparseblossom} that we use in \Cref{sec:results}. 
The detailed results of these simulations are shown in \Cref{app:cultivation}.

\Cref{tab:cultivation} shows results from these simulations for $p = 10^{-3}$, the highest error rate we consider.
Code distance $d=7$ is the most constricted scenario, where
the number of code cycles of a complete cultivation is $r=9$.
The effective success rate is $0.49$, indicating the average duration of a successful cultivation is $ r_\text{eff} = r/0.49 \approx 2r = 18 \approx 2.5 d$ (for $d=7$).
$r_\text{eff} < 2.5d$ for all other choices of $(d,p)$ considered in this work.
We note that the decoding assumption can be relaxed by tuning the complementary gap threshold to allow a higher success rate. The cultivation error rate will increase correspondingly, but it does not impact our pipeline or main conclusions.
Moreover, there is evidence that cultivation can be further improved~\cite{hirano2025efficient}.

\subsection{Clifford $\to$ lattice surgery}
\label{sec:lattice-surgery-compilation}
We merge the Clifford gates from the $T$ sequence normalization with those originally in the circuit to form Clifford subcircuits, separated by the $T_XT_Z$ sequences. We employ \texttt{LaSsynth}~\cite{lassynth} to synthesize the lattice surgery operations that implement the logic of each subcircuit. The logic is described by stabilizer flows~\cite{aaronson2004improvedsimulation}, a gateless representation of Clifford circuits, which we extract using \texttt{Stim}~\cite{gidney2021stim}. \texttt{LaSsynth} encodes the circuit logic, spatial constraints, and lattice surgery validity into a SAT problem, which can be solved by widely studied SAT solvers.  Contrary to past studies, there is no distinction between data and ancilla qubits, besides that at the end of the lattice surgery, we require the qubit of the subsequent $T_XT_Z$ sequence to be placed at the center of the round-robin injection, and that the other data qubits do not interfere with the cultivation qubits. This flexibility, along with the exhaustiveness of the SAT-solver-based construction, enables constructions with modest spacetime cost.
We then compose the lattice surgery operations from the Clifford subcircuits and the $T_XT_Z$ sequences to assemble the full lattice surgery corresponding to the Clifford+$T$ circuit.

\subsection{Lattice surgery $\to$ physical circuit}
\label{sec:physical-circuit}
Converting lattice surgery to the physical circuit is conceptually straightforward, as temporal evolution, merging, and splitting are all implemented with well-defined physical circuits (for a given code distance).
Nonetheless, composing physical subcircuits together coherently while maintaining fault tolerance and code distance is an involved process.
To automate and ensure the correctness of this process, we make use of and improve some of the features of the design automation tool \texttt{tqec}~\cite{tqec}.
We note that our compiled physical circuits are ``Clifford proxy'' circuits. In particular, they do not include the magic state cultivation protocols, and we replace magic state ($\ket{T}$) inputs with $Z$-basis ($\ket{0}$) initializations\footnote{$Y$-basis initializations and measurements are still under development in \texttt{tqec}, so we also approximate $S$ gates using $Z$ gates at this moment.}. Due to the current limitations of \texttt{tqec} and \texttt{Stim}, the circuits also do not perform conditional $Y$-basis measurements for the injections.

Finally, in order to obtain a logical measurement outcome from the execution of an encoded quantum algorithm, we also need to identify how physical measurement outcomes are combined to recover the target logical observables.
This problem reduces to that of identifying \emph{correlation surfaces}~\cite{raussendorf2007fault, fowler2008topological} in a pipe diagram.
Computing correlation surfaces is a necessary step in surface code compilation and has long been neglected by past compilation works due to the absence of physical-level compilation.
\texttt{tqec}'s algorithm for computing correlation surface scales exponentially in the number of lattice surgery merges and is intractable for even lattice surgery representations compiled from small circuits.
We develop a polynomial algorithm to support our algorithm-scale compilations, detailed in~\cite{tianyi_cs}.

\section{Compilation and Algorithmic Break-Even}
\label{sec:results}

\subsection{Target applications}
We showcase two representative applications of our compilation pipeline that are suitable for near-term demonstrations: quantum phase estimation (QPE) and the quantum approximate optimization algorithm (QAOA).
We then simulate circuits produced by our pipeline to predict the hardware resources and specifications required to achieve algorithmic break-even for QAOA.

\subsubsection{Quantum Phase Estimation}
QPE~\cite{kitaev1995quantum} is a common algorithmic subroutine that estimates the eigenvalue of a given eigenstate of a given unitary operator (note that this eigenvalue is a phase, $e^{i\theta}$, hence the name QPE).
QPE is not only indispensable in estimating the ground-state energy in quantum simulation, but also a cornerstone powering many other quantum algorithms, such as Shor's algorithm and the HHL algorithm.
QPE is nontrivial to implement, with a circuit structure that demands non-local connectivity and controlled unitaries.
We compile QPE with 4-bit precision for the unitary $R_Z(\pi/4)$.

\begin{table*}[ht]
\caption{
Characteristics of the compiled results.
Here spacetime volume is the number of cubes in a pipe diagram.
The total number of physical qubits (including data, syndrome, and padding qubits) used by circuits compiled for $d=(7,9,11)$ are $(1101, 1737, 2517)$.
}
{\small
\begin{tabular}{c  ccc ccc ccc  ccc ccc ccc}
\toprule
~ & \multicolumn{9}{c}{QPE ($R_Z(\pi/4)$, 4-bit precision)} & \multicolumn{9}{c}{QAOA (5-bit LABS)} \\
\cmidrule(lr){2-10}\cmidrule(lr){11-19}
~ & \multicolumn{3}{c}{Spacetime volume} & \multicolumn{3}{c}{Code Cycles, $n_\mathrm{cycles}$} & \multicolumn{3}{c}{$T$ count} & \multicolumn{3}{c}{Spacetime volume} & \multicolumn{3}{c}{Code Cycles, $n_\mathrm{cycles}$} & \multicolumn{3}{c}{$T$ count} \\
\cmidrule(lr){1-1}\cmidrule(lr){2-4}\cmidrule(lr){5-7}\cmidrule(lr){8-10}\cmidrule(lr){11-13}\cmidrule(lr){14-16}\cmidrule(lr){17-19}
Distance, $d$ & 7 & 9 & 11 & 7 & 9 & 11 & 7 & 9 & 11 & 7 & 9 & 11 & 7 & 9 & 11 & 7 & 9 & 11\\
\cmidrule(lr){1-1}\cmidrule(lr){2-4}\cmidrule(lr){5-7}\cmidrule(lr){8-10}\cmidrule(lr){11-13}\cmidrule(lr){14-16}\cmidrule(lr){17-19}
$p=10^{-3}$ & 1210 & 1950 & 2403 & 1148 & 2349 & 3476 & 69 & 151 & 205 & 726 & 1166 & 1811 & 693 & 1395 & 2596 & 42 & 87 & 172 \\
$p=7\times10^{-4}$ & 1950 & 2391 & 2687 & 1827 & 2835 & 3872 & 151 & 205 & 234 & 1172 & 1739 & 2212 & 1085 & 2043 & 3135 & 87 & 172 & 226 \\
$p=5\times10^{-4}$ & 2401 & 2700 & 3523 & 2212 & 3186 & 5027 & 205 & 240 & 343 & 1739 & 2288 & 2577 & 1589 & 2655 & 3641 & 172 & 239 & 273 \\
\bottomrule
\end{tabular}
}
\label{tab:pipe_line_data}
\end{table*}

\subsubsection{QAOA and the LABS problem}

The quantum approximate optimization algorithm (QAOA)~\cite{Hogg2000,Hogg2000search,farhi_qaoa} solves combinatorial optimization problems by applying a parameterized quantum circuit to a fixed initial state. The parameters in the circuit are chosen such that sampling the resulting state gives high-quality solutions with high probability. 
The QAOA state after $l$ layers is given by:
\begin{equation}
|\beta, \gamma\rangle = \prod_{j=1}^{l} e^{-i \beta_jH_M} e^{-i \gamma_j H_P} \ket{+}^{\otimes N} ,
\label{eq:QAOA}
\end{equation}
where $\bm\gamma, \bm \beta$ are free parameters, $H_P$ is a Hamiltonian that encodes an objective function, and $H_M$ is a mixing Hamiltonian that we set to be $H_M=\sum_j X_j$, where $X_j$ denotes the tbit-flip operator for qubit $j$.
QAOA has a similar circuit structure to Hamiltonian simulation and variational quantum algorithms, and has been a candidate for demonstrating algorithmic break-even in EFT architectures~\cite{he2025performance, jin2025iceberg}.

We apply QAOA to the low autocorrelation binary sequences (LABS) problem~\cite{Boehmer1967,Schroeder1970}, which has applications in communications engineering~\cite{Boehmer1967,Golay1977}. Solving LABS with QAOA is a promising candidate for quantum advantage in optimization since LABS becomes classically intractable at $\approx100$ binary variables~\cite{OPUS2-git_labs-Boskovic,Packebusch2016} and since QAOA offers a polynomial speedup over classical state-of-the-art for LABS~\cite{shaydulin2024evidencelabs}.
$N$-bit LABS is encoded on qubits by the Hamiltonian:
\begin{align}
\label{eq:labs_energy}
  H_P(N) = \sum_{k=1}^{N-1} A_k^2,
  &&
  \text{where}
  &&
  A_k = \sum_{i=1}^{N-k} Z_i Z_{i+k}.
\end{align}
We remark that for the LABS problem, Ref.~\cite{shaydulin2024evidencelabs} provides preoptimized QAOA parameters for small $N$ and instance-independent parameters for large $N$, removing the need for variational parameter optimization.
We evaluate our pipeline by compiling QAOA with $p=1$ applied to 5-bit LABS, whose Hamiltonian is 
\begin{align}
    H_P(5) & = 2\left(Z_0Z_1Z_3Z_4+Z_1Z_2Z_3Z_4+Z_0Z_1Z_2Z_3\right) \notag \\
    &\qquad + Z_0Z_2 + Z_0Z_4 + Z_1Z_3 + Z_2Z_4.
\end{align}

\subsection{Results}

\begin{figure}
    \centering
    \includegraphics[width=\linewidth]{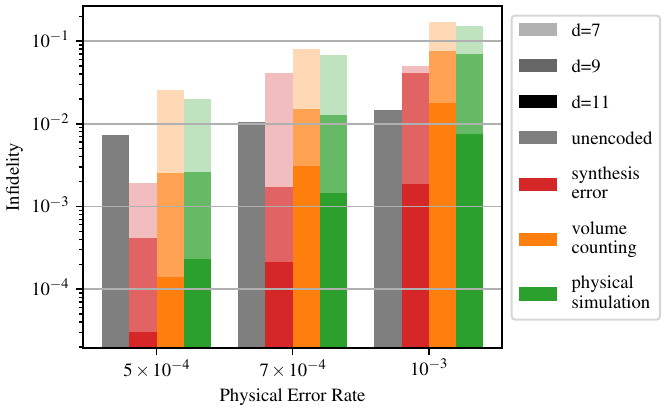}
    \caption{Comparison of infidelity estimation by logical-level simulation, spacetime volume counting, and physical-level simulation.}
    \label{fig:comparison}
\end{figure}

\begin{figure}
    \centering
    \includegraphics[width=\linewidth]{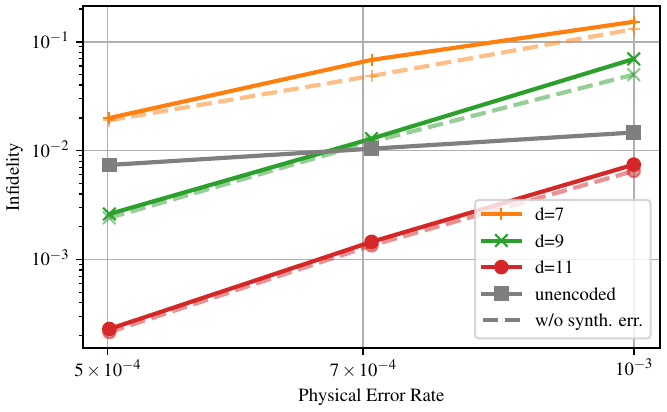}
    \caption{Performance of running the algorithm with the surface code versus directly, unencoded. Dashed lines indicate sampling error $\varepsilon$ obtained from the proxy Clifford simulations, without synthesis error.}
    \label{fig:infidelity_plot}
\end{figure}

With the compiled physical circuits, our pipeline provides more accurate physical-level information than previous works on resource estimation.
These circuits are also compiled adaptively to each choice of code distance $d$ and physical error rate $p$ at the rotation synthesis step, with the aim of minimizing the overall error.
We display characteristics of the compiled physical circuits in \Cref{tab:pipe_line_data}. 

The Clifford proxy circuits produced by the pipeline are efficiently simulable on a classical computer. 
Subjecting the circuits to the SI1000 noise model~\cite{Gidney_SI1000}, we sample 500,000 shots of the circuit using \texttt{Stim} and \texttt{PyMatching}~\cite{pymatching}.
This sampling yields an error rate $\epsilon$ at which logical observables are flipped (erroneously) due to physical errors, which is approximately equal to the infidelity of the executed circuit.
We combine execution fidelity with the synthesis fidelity $\mathcal{F}_\text{synth} =\lvert\braket{\Psi_L|\Psi_T}\rvert^2$, where $\ket{\Psi_L}$ and $\ket{\Psi_T}$ are, respectively, the states produced by the (undecomposed) logical and (synthesized) Clifford+$T$ circuits.
The overall fidelity $\mathcal{F}_\mathrm{combined}$ of the algorithm is then approximately the product of execution, synthesis, and cultivation fidelities, $\mathcal{F}_\mathrm{combined} = (1-\varepsilon)\times\mathcal{F}_\text{synth}\times(1-e_\mathrm{cultiv})^{\#T}$, where $e_\mathrm{cultiv}$ is the cultivation error rate obtained in \Cref{sec:t-injection} and $\#T$ is the $T$ gate count.

We also simulate the unencoded version of the logical circuits after compiling them to the same hardware constraints, such as 2D nearest-neighbor connectivity.
We perform density matrix simulation with the same SI1000 noise model and compute the fidelity $\mathcal{F}_\mathrm{physical} = \braket{\Psi_L|\rho_{\mathrm{noisy}}|\Psi_L}$, where $\rho_{\mathrm{noisy}}$ represents the density matrix of the noisy system.
Finally, we compute an alternate estimation of the execution fidelity by counting the number of cubes $V$ in a pipe diagram and multiply it by the synthesis and cultivation fidelities, $\mathcal{F}_\text{vol} = (1 - e_\mathrm{unit})^{V}\times\mathcal{F}_\text{synth}\times(1-e_\mathrm{cultiv})^{\#T}$, where $e_\mathrm{unit}$ is the logical error rate for a unit spacetime volume. 

We summarize infidelities for QAOA circuits compiled to code distances $d=7, 9, 11$ and physical error rates $p=5\times 10^{-4}, 7 \times 10^{-4},$ and $10^{-3}$ in \Cref{fig:comparison}, where an the infidelity $\mathcal{I}$ corresponding to fidelity $\mathcal{F}$ is $\mathcal{I} = 1-\mathcal{F}$.
We present the infidelities for unencoded execution and compiled physical circuits (a subset of the data in \Cref{fig:comparison}) in \Cref{fig:infidelity_plot}, allowing us to compare the performance of running the algorithm fault-tolerantly versus without QEC.
We estimate the break-even to occur at $d=11$ ($2517$ qubits) for $p=10^{-3}$ and $d=9$ ($1737$ qubits) for $p=5\times 10^{-4}$.
The results are similar for QPE: for $p=10^{-3}$ break-even occurs at $d=11$ with $\mathcal{I} = 2.71 \times 10^{-2}$, compared with $\mathcal{I}_\text{unencoded} = 4.89 \times 10^{-2}$.
For $p=5\times 10^{-4}$ break-even occurs at $d=9$ with $\mathcal{I} = 3.50 \times 10^{-3}$, compared with $\mathcal{I}_\text{unencoded} = 2.49 \times 10^{-2}$.

\section{Conclusion}
\label{sec:outlook}

In this work, we demonstrate a pipeline for compilation of logical circuits into physical circuit based on 2D surface code and lattice surgery. Our pipeline allows us to identify algorithmic break-even conditions by performing complete physical circuit simulations of a Clifford proxy of the logical circuit, altogether simulating an algorithm running on thousands of physical qubits.

We demonstrate our pipeline by compiling QPE and QAOA with $5$ logical qubits supported on $9$ surface code patches, and making concrete predictions for the requirements to achieve algorithmic break-even for the logical QAOA circuit.
Looking forward, tackling algorithms with tens of logical qubits will require modifying our use of \texttt{LaSsynth}~\cite{lassynth} with a scalable approach to optimizing lattice surgery operations, for example, by performing ``peephole'' optimizations of the pipe diagram.
Our pipeline can also be augmented to convert pipe diagrams into dynamic programs that can accommodate nondeterministic magic state preparation routines, thereby producing a complete representation of the physical instructions that execute a logical algorithm on hardware.

Our results show that surface code with distance $\geq 9$ is sufficient to run nontrivial algorithms beyond break-even.
We challenge the community to demonstrate break-even with these circuits on near-term hardware.

\section*{Acknowledgments}

The authors thank Rob Otter for the executive support of the work and invaluable feedback on this project. The authors appreciate the assistance from Bochen Tan and the \texttt{tqec} community, especially Yiming Zhang, for using and adapting \texttt{LaSsynth} and \texttt{tqec}. The authors thank Ryan Babbush, Tanuj Khattar, and other members of the Google Quantum AI team for providing helpful feedback on a draft version of this work.

\bibliographystyle{ACM-Reference-Format}
\bibliography{ref.bib}

\section*{Disclaimer}
This paper was prepared for informational purposes by the Global Technology Applied Research center of JPMorganChase. This paper is not a product of the Research Department of JPMorganChase or its affiliates. Neither JPMorganChase nor any of its affiliates makes any explicit or implied representation or warranty and none of them accept any liability in connection with this position paper, including, without limitation, with respect to the completeness, accuracy, or reliability of the information contained herein and the potential legal, compliance, tax, or accounting effects thereof. This document is not intended as investment research or investment advice, or as a recommendation, offer, or solicitation for the purchase or sale of any security, financial instrument, financial product or service, or to be used in any way for evaluating the merits of participating in any transaction.

\appendix
\onecolumn

\section{Circuit Details}

Here we share more details about the circuits that we simulated to obtain the results in \Cref{fig:comparison,fig:infidelity_plot}.
For all circuit simulations we use SI1000, a superconducting-inspired noise model~\cite{Gidney_SI1000} parameterized by a two-qubit error rate $p$:

\begin{table}[h]
\caption{
SI1000 noise model.
}
{\small
\begin{tabular}{cccccc}
\toprule
Two-qubit Cifford & Single-qubit Clifford & Measurement Error & Reset Error & Idling Error & Resonator Idling\\
\cmidrule(lr){1-1}\cmidrule(lr){2-2} \cmidrule(lr){3-3} \cmidrule(lr){4-4} \cmidrule(lr){5-5} \cmidrule(lr){6-6}
$p$ & $p/10$ & $5p$ & $2p$ & $p/10$ & $2p$ \\
\bottomrule
\end{tabular}
}
\label{tab:cultivation}
\end{table}

\begin{figure}[htbp]
    \begin{subfigure}[b]{0.9\linewidth}
        \begin{tabular}{l}
        \includegraphics[height = 1.8cm]{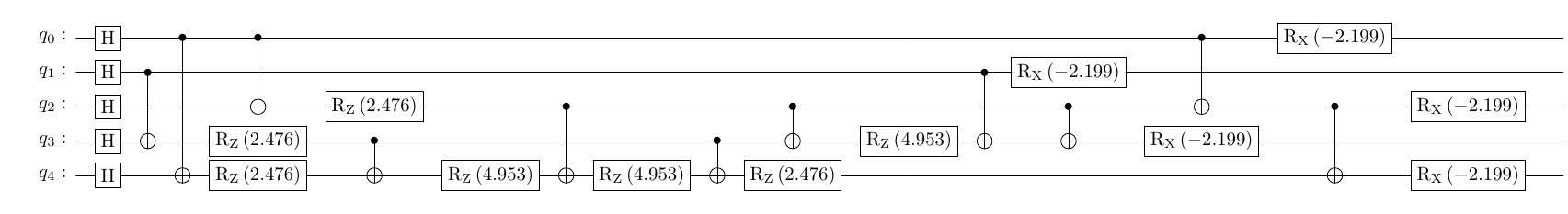} 
        \end{tabular}
        \caption{}
        \label{fig:circuit1}
    \end{subfigure}
    \hfill
    \begin{subfigure}[b]{0.9\linewidth}
        \begin{tabular}{l}
        \includegraphics[height = 1.8cm]{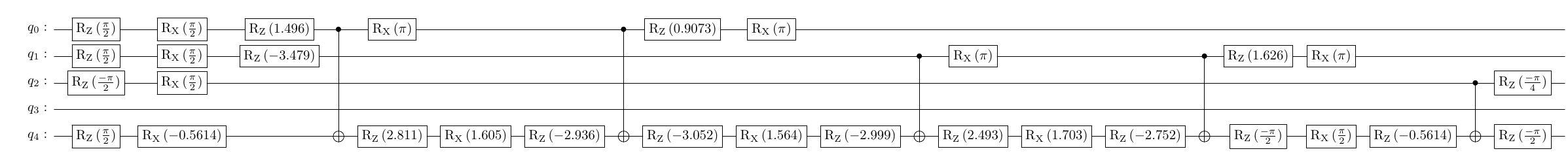} \\
        \includegraphics[height = 1.8cm]{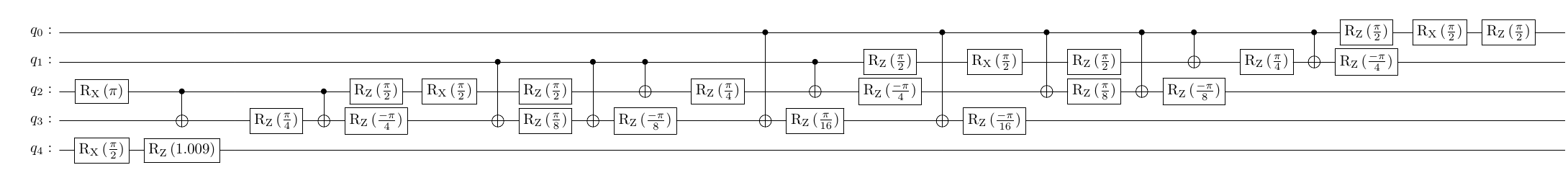}
        \end{tabular}
        \caption{}
        \label{fig:circuit2}
    \end{subfigure}
    \caption{(a) Logical QAOA circuit for 5-bit LABS. (b) QPE logical circuit for 4-bit precision $R_Z(\pi/4)$.}
    \label{fig:logical_circuits}
\end{figure}

\begin{figure}[h]
  \centering
  \begin{tabular}{l}
  \includegraphics[scale=0.65]
  {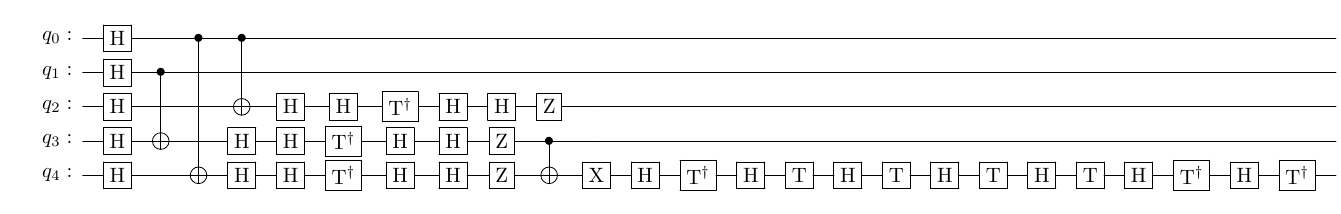}
  \\
  \includegraphics[scale=0.65]
  {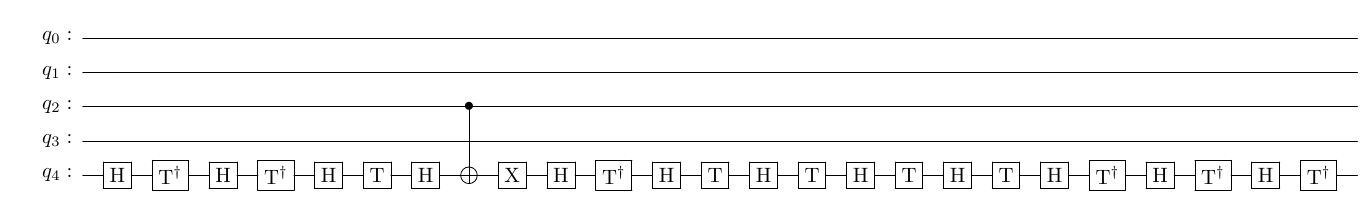}
  \\
  \includegraphics[scale=0.65]
  {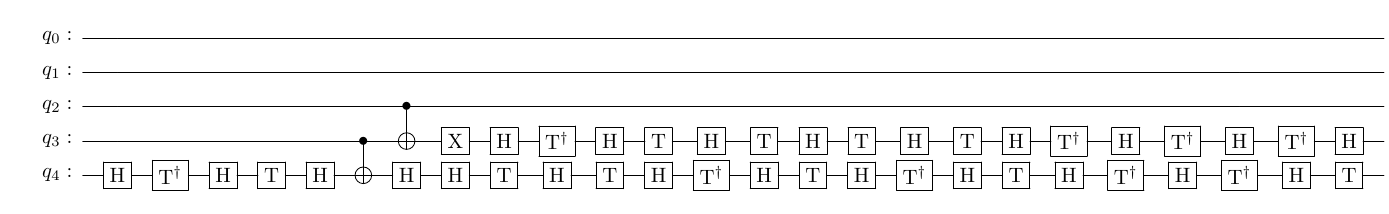}
  \\
  \includegraphics[scale=0.65]
  {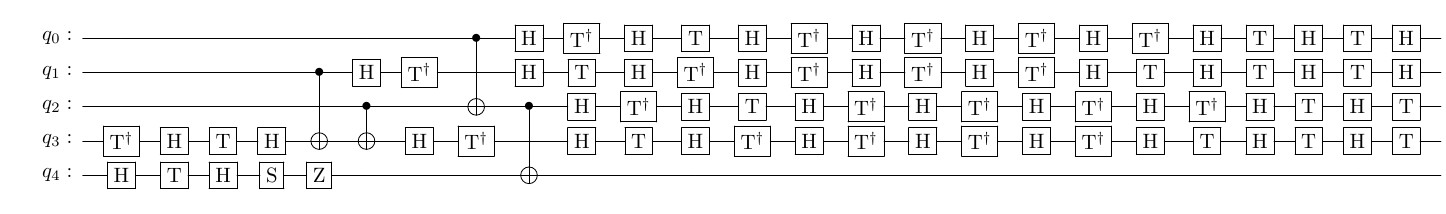}
  \\
  \includegraphics[scale=0.65]
  {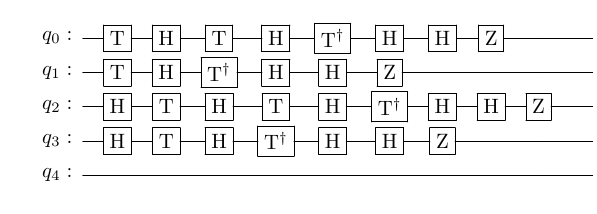}
  \end{tabular}
  \caption{The synthesized Cliffort+$T$ circuit for 5-bit LABS at code distance $d=9$ and physical error rate $p=10^{-3}$. Note $HTH = T_X$.}
  \label{fig:ft_circuit}
\end{figure}

The Clifford + $R_Z$ circuits for the selected algorithms shown in \Cref{fig:logical_circuits} cannot be run fault-tolerantly on the surface code. To convert them to fault-tolerant circuits, like the example shown in \Cref{fig:ft_circuit}, we must synthesize the single-qubit rotations into a discrete gate set. As described in the text, we do so in an error-aware way, factoring logical gate errors and idling errors into the $U3\to\mathrm{Clifford}+T$ synthesis protocol to obtain an optimal sequence length. To model the noise of the $T$ gates, we combine two numerically obtained error rates. Following \cite{gidney2024magicstatecultivationgrowing}, we perform cultivation of $Y$-states for each $(d,p)$ (see \Cref{app:cultivation}) and then multiply the resulting logical error rate by 2 to estimate the error $e_\mathrm{cultiv}$ of cultivating a $T$-state. During the injection process (see \Cref{fig:full-pipeline}(d)), each of the data qubits is occupied for $d$ rounds, and the ancilla is occupied, on average, for $1.25d$ rounds. We estimate the logical T-gate infidelity by 

\begin{equation}
    e_{T} = 1 - (1 - e_\mathrm{unit})^{n+1.25}(1 - e_\mathrm{cultiv}), 
\end{equation}
where for each $(d,p)$ the unit error rate $e_\mathrm{unit}$ is obtained via a memory experiment. The error $e_{T}$ is then used as input for the rotation synthesis.

\newpage

\section{Cultivation}
\label{app:cultivation}
\begin{figure}[htbp]
    \centering
    \begin{subfigure}[b]{0.85\linewidth}
        \centering
        \includegraphics[width=\textwidth]{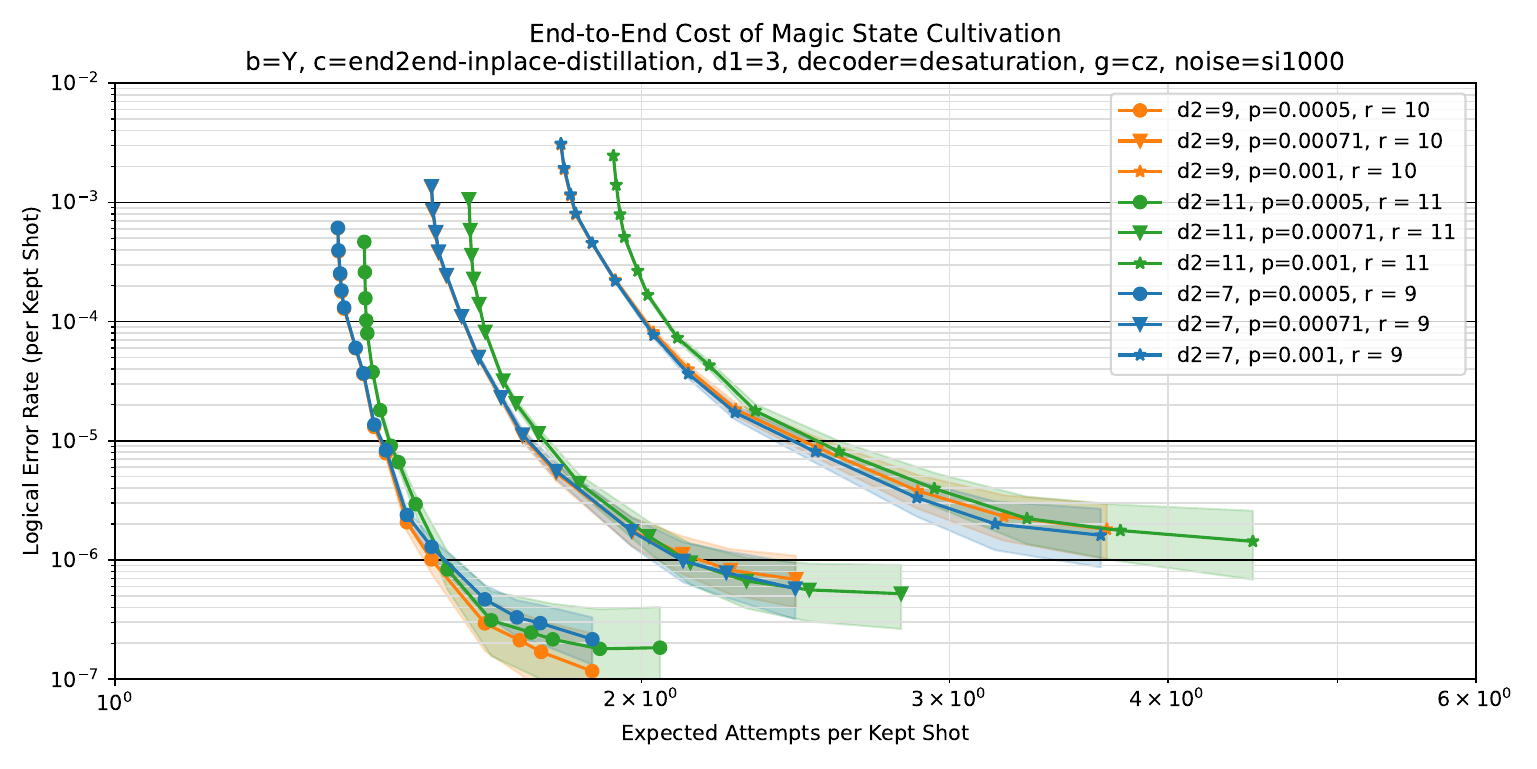}
        \caption{}
        \label{fig:cultivation_plot1}
    \end{subfigure}
    \hfill
    \begin{subfigure}[b]{0.9\linewidth}
        \centering
        \includegraphics[width=\textwidth]{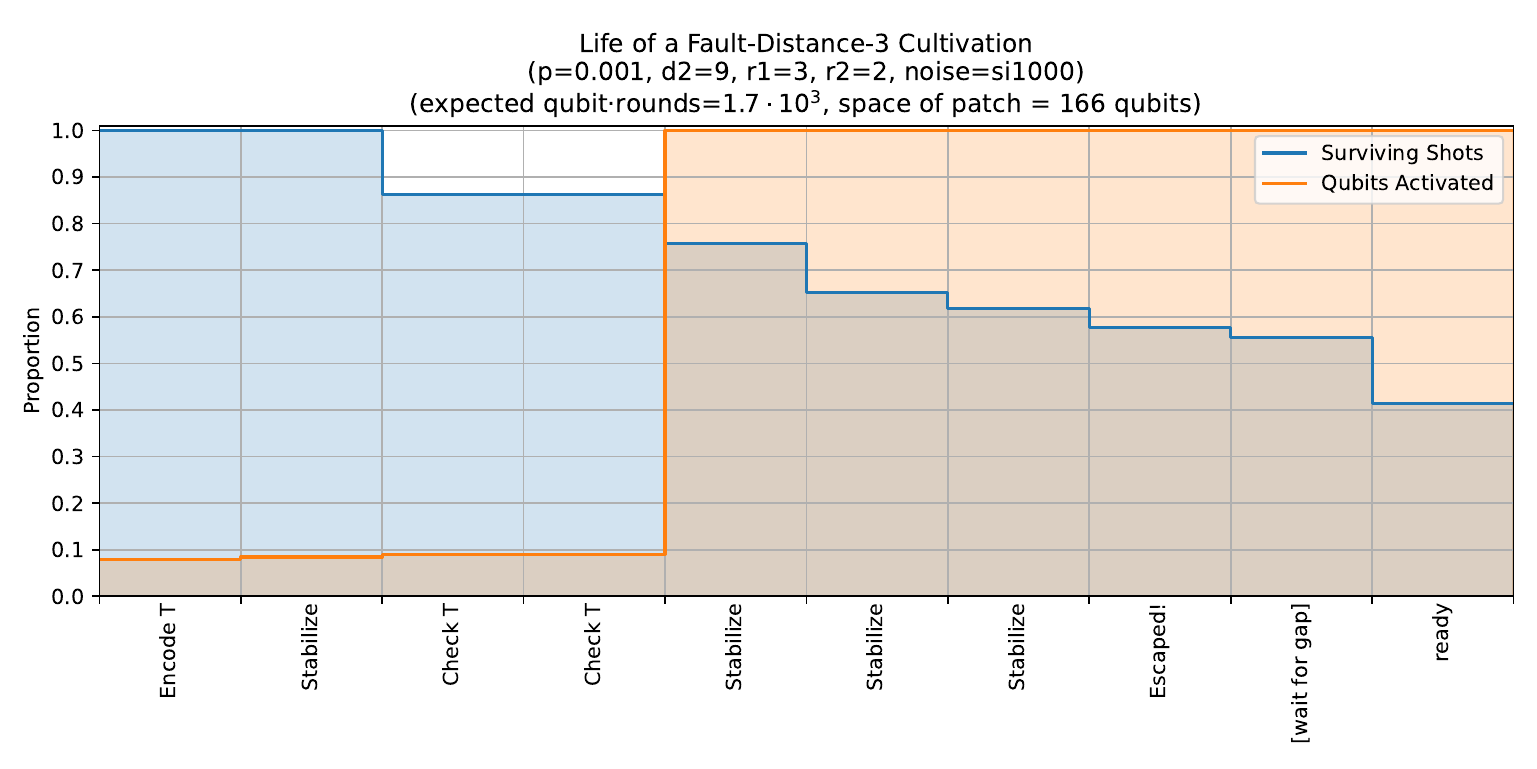}
        \caption{}
        \label{fig:cultivation_plot2}
    \end{subfigure}
    \caption{ (a) Logical error rate vs Expected attempts per kept shot for each $(d,p)$. Note that the expected attempts per kept shot does not factor into account that the color code cultivation can occur multiple times in parallel in the space allocated to the final code. (b)
    Surviving shots and qubit use at each stage of the cultivation protocol for $(d=9, p = 10^{-3})$.}
    \label{fig:cultivation_plot}
\end{figure}

Magic state cultivation is one of the key components used in our pipeline. As discussed in the text, the round-robin injection scheme requires that the cultivation protocol finish in $2.5d$ rounds. By lowering the threshold value of the complementary gap, we can decrease the expected wait time of a successful cultivation run at the expense of a worse logical error rate for the process. This relationship is shown in \Cref{fig:cultivation_plot}(a), where we plot the logical error rate versus the expected attempts per kept shot for each $(d,p)$. On a given curve, each point represents a different threshold complementary gap, with stringency increasing to the right. 

In \Cref{fig:cultivation_plot}(b), the number of surviving shots at the rightmost bin gives the x-axis values (expected attempts per kept shot) in \Cref{fig:cultivation_plot}(a).
As an example, in the specific case of $(d,p)=(9, 10^{-3})$, the part of the process involving a $d=3$ color code succeeds $86\%$ of the time and can be implemented multiple times in parallel in the space allocated for the final code. Therefore, we use the effective success rate $.41/.86 = .477$ to estimate the expected time per successful cultivation and check whether the particular choice of complementary gap yields a cultivation protocol that fits into $2.5d$ rounds. An analogous effective success rate is used for the other data points $(d,p)$.

\end{document}